\documentclass[pra,10pt,twocolumn,superscriptaddress,showpacs]{revtex4}
\usepackage{amsmath}
\usepackage{latexsym}
\usepackage{amssymb}
\usepackage{graphics,epstopdf}
\usepackage{graphicx}
\usepackage[colorlinks=true, citecolor=blue, urlcolor=blue ]{hyperref}
\usepackage{float}
\usepackage{graphicx}
\usepackage{amsfonts}
\usepackage{appendix}
\begin{document}

\title{Wigner's form of the Leggett-Garg inequality, No-Signalling in Time, and Unsharp Measurements}

\author{Debashis Saha}
\email{debashis7112@iiserkol.ac.in}
\affiliation{Indian Institute of Science Education and Research Kolkata, Mohanpur 741252, India}

\author{Shiladitya Mal}
\email{shiladitya@bose.res.in}
\affiliation{Satyendra Nath Bose National Centre for Basic Sciences, Kolkata 700098, India}

\author{Prasanta K. Panigrahi}
\email{pprasanta@iiserkol.ac.in}
\affiliation{Indian Institute of Science Education and Research Kolkata, Mohanpur 741252, India}

\author{Dipankar Home}
\email{quantumhome80@gmail.com}
\affiliation{Center for Astroparticle Physics and Space Science (CAPSS), Bose Institute, Kolkata 700091, India}

\begin{abstract}
Wigner's form of the local realist inequality is used to derive its temporal version for an oscillating two-level system involving two-time joint probabilities. Such an inequality may be regarded as a novel form of the Leggett-Garg inequality (LGI) constituting a necessary condition for macrorealism. The robustness of its quantum mechanical (QM) violation against unsharpness of measurement is investigated by using a suitable model of unsharp measurements. It is found that there exists a range of values of the sharpness parameter (characterizing precision of the relevant measurements) for which the usual LGI is satisfied by QM, but Wigner's form of the LGI  (WLGI)  is violated. This implies that for such unsharp measurements, the QM violation of macrorealism cannot be tested using the usual LGI, but can be tested using WLGI. In showing this, we take into account the general form of the usual LGI involving an arbitrary number of pairs of two-time correlation functions. Another recently proposed necessary condition for macrorealism, called `no-signalling in time', is also probed, showing that its QM violation persists for arbitrarily unsharp measurements.
\end{abstract} 


\maketitle

\section{Introduction} 
Complementing the exploration of the nonclassical features of the microphysical world, investigation of the fundamental aspects and validity of quantum mechanics (QM) at the macroscopic level has been attracting an increasing attention over the past decade \cite{ex1,ex2,ex3,ex4,ex5,jor, ex7,ex8,LGI2}. For this latter line of study, a key ingredient is provided by the Leggett-Garg form of macrorealist inequality (LGI) \cite{LGI1,LGI3} which is a temporal analogue of Bell's inequality involving testable temporal correlation functions, and whose validity can be considered a necessary condition for what is regarded as macrorealism. LGI is derived from the notion of macrorealism which is characterized by the following assumptions - {\it Macrorealism per se:} At any given instant, irrespective of any measurement, a macroscopic object is in one of the available states, having definite values for all its observable properties. {\it Non-invasive measurability:} It is possible, at least in principle, to determine which of the states a system is in, without affecting the state itself or the system's subsequent behavior. \\
 
A wide range of studies have probed aspects of LGI and its QM violation for a variety of systems; see, for example, a recent comprehensive review \cite{review}. Against this backdrop, an earlier unexplored variant of LGI is formulated in the present paper by developing an analogy to the argument used in deriving Wigner's form of the local realist inequality \cite{Wigner}. Here we invoke, instead of local realism, the assumption of macrorealism in the context of an oscillating two-level system. This is parallel to the way the standard form of LGI \cite{LGI1, LGI3} involving two-time correlation functions is obtained in analogy to the Bell-CHSH local realist inequality \cite{bell, CHSH}. The quantum mechanical (QM) violation of thus obtained, what may be called, Wigner's form of the LGI (WLGI) is studied by considering, for simplicity, the two-state oscillation pertaining to a system oscillating between the degenerate orthogonal eigenstates of a Hamiltonian where the oscillation is induced by an external field. Also, to be noted that this type of example is relevant to some of the actual experiments probing LGI \cite{review}. For such an example, the robustness of the QM violation of WLGI with respect to imprecision of the relevant measurements is investigated in this paper by using the formalism of what is known as unsharp measurement \cite{busch,PB, busch0, spek, busch1}. Interestingly, it is found that there exists a range of values of the sharpness parameter characterizing precision of the relevant measurements for which QM satisfies the usual form of LGI, but violates WLGI, thereby signifying the non-equivalence between LGI and WLGI. In particular, the QM violation of WLGI persists for smaller values of the sharpness parameter (that correspond to greater imprecision of the measurements involved) compared to that for the usual form of LGI. Hence, for such unsharp measurements, the QM violation of macrorealism can be shown using WLGI, but not in terms of LGI. This result is, thus, of interest in the context of the question as regards the extent of imprecision or unsharpness of the relevant measurements for which the QM violation of macrorealism is manifested. Here it needs to be noted that while making the required comparison between WLGI and LGI, we take into account the general form of the usual LGI involving $n$ pairs of two-time correlation functions. It is then found that for the situations studied in this paper, pertaining to an oscillating two-level system, the QM violation of WLGI is more robust than that of LGI in the presence of unsharp measurements of the type considered here.

Furthermore, we consider the other proposed necessary condition for macro-realism known as `No-Signaling in Time' (NSIT) suggested by Kofler and Brukner \cite{nsit} which stipulates that the statistics of the outcomes of measurements at any instant should \textit{not} show dependence on whether any prior measurement has been performed; in other words, NSIT is based on applying the condition of \textit{non-invasive measurability} (NIM) at the statistical level whose violation would imply violation of NIM at the individual level. Here we may note that Leggett \cite{LGI1, LGI3} has argued that NIM can be regarded to be a `natural' corollary of the condition of \textit{macrorealism per se} in the context of what is known as an ideal negative result measurement (applicable for testing both LGI and WLGI) that can, too, be invoked for testing NSIT, as has been noted by Kofler and Brukner \cite{nsit}. By investigating the robustness of the QM violation of NSIT against unsharpness of measurement, a striking result is obtained that, no matter whatever be the value of the sharpness parameter characterizing precision of the relevant measurements, QM violates NSIT. In other words, the QM violation of NSIT turns out to be most robust against unsharp measurement of the type considered in the present paper.

\section{Wigner's form of LGI}

We begin by recapitulating Wigner's original argument \cite{Wigner} that derived a local realist testable inequality for a pair of spatially separated spin-1/2 particles in the singlet state. This was based on  assuming as a consequence of local realism, the existence of overall joint probabilities of the definite  outcomes of measuring the relevant dichotomic observables of the two particles that would yield the pair-wise marginal joint probabilities which are actually measurable. In the scenario studied by Wigner, the spin components of each of the two spatially separated particles are taken to be measured along three respective directions, say, $\hat{a}, \hat{b}$ and $\hat{c}$. Then consider, for example, the observable joint probability of obtaining both the outcomes +1 if, say, $\overrightarrow{\sigma}.\hat{a}$ and $\overrightarrow{\sigma}.\hat{b}$ are measured on the first and the second particle respectively, denoted as $P(\hat{a}+ , \hat{b}+)$. Using the perfect anticorrelation property of the singlet state in question, $P(\hat{a}+ , \hat{b}+)$ can be written as a marginal in the form $P(\hat{a}+ , \hat{b}+) = \rho(+, -, + ; - , +, -) + \rho(+, -, -; -, +, +)$ with $\rho(v_{1}(\hat{a}), v_{1}(\hat{b}) , v_{1}(\hat{c}); v_{2}(\hat{a}), v_{2}(\hat{b}), v_{2}(\hat{c}))$ to be the overall joint probability of the definite outcomes of measurements pertaining to all the relevant observables, where $v_1(\hat{a})$ represents the outcome ($\pm1$) of measurement of the observable $\hat{a}$ for the first particle, and so on. Similarly, considering the expressions for the observable joint probabilities $P(\hat{c}+, \hat{b}+)$ and $P(\hat{a}+, \hat{c}+)$ as marginals, and assuming non-negativity of the overall joint probability distributions, it follows that 

\begin{equation}
\label{wig}
P(\hat{a}+ , \hat{b}+) - P(\hat{a}+, \hat{c}+) - P(\hat{c}+, \hat{b}+) \leq 0
\end{equation}

\noindent which is one of the various forms of Wigner's version of the local realist inequality.\\

Next, in order to obtain the temporal version of Wigner's inequality (\ref{wig}) by developing an appropriate analogy with the preceding argument,  we proceed as follows. Let us focus our attention on an ensemble of systems undergoing temporal evolution involving oscillation between the two states, say, 1 and 2, and let $Q(t)$ be an observable quantity such  that, whenever measured, it is found to take a value +1(-1) depending on whether the system is in the state 1(2). Now, consider a collection of sets of experimental runs, with each set of runs starting from the same initial state. On the first set of runs, let $Q$ be measured at times $t_1$ and $t_2$, on the second $Q$ be measured at $t_2$ and $t_3$, and on the third at $t_1$ and $t_3$ (here $t_1<t_2<t_3$). From such measurements one can then determine the pair-wise joint probabilities like $P(Q_1, Q_2), P(Q_2, Q_3),P(Q_1, Q_3)$ where $Q_i$ is the outcome ($\pm 1$) of measuring $Q$ at $t_i$, $i = 1, 2, 3$. In this scenario, it is possible to suitably adapt the argument leading to Wigner's inequality (\ref{wig}) with the times $t_i$ of measurement playing the role of apparatus settings. Here we note that, as a consequence of the assumption of \textit{macrorealism per se}, one can infer the existence of overall joint probabilities $\rho(Q_1, Q_2, Q_3)$ pertaining to different combination of outcomes for the relevant measurements, while the assumption of \textit{non-invasive measurability} implies that such overall joint probabilities would remain unaffected by measurements, and hence, by appropriate marginalization, the pair-wise observable joint probabilities can be obtained. For example, the observable joint probability $P(Q_{2}+, Q_{3}-)$ of obtaining the outcomes +1 and $-1$ for the sequential measurements of $Q$ at the instants $t_2$ and $t_3$ respectively can be written as 
\begin{eqnarray} 
\label{eq2}
\begin{split}
& P(Q_{2}+,Q_{3}-)=\sum_{Q_{1}=\pm}\rho(Q_{1},+,-)\\&
=\rho(+,+,-)+\rho(-,+,-)
\end{split}
\end{eqnarray}
Writing similar expressions for the other measurable marginal joint probabilities $P(Q_{1}-,Q_{3}-)$ and $P(Q_{1}+,Q_{2}+)$, we get
\begin{equation}
\label{eq3}
\begin{split}
&P(Q_{1}+,Q_{2}+)+ P(Q_{1}-,Q_{3}-) - P(Q_{2}+,Q_{3}-) \\& = \rho(+,+,+)+\rho(-,-,-)
\end{split}
\end{equation}
Then, invoking non-negativity of the joint probabilities occurring on the RHS of Eq(\ref{eq3}), the following form of WLGI is obtained in terms of three pairs of two-time joint probabilities
\begin{equation}
\label{wlgia}
P(Q_2+,Q_3-) - P(Q_1+,Q_2+) - P(Q_1-,Q_3-) \leq 0   
\end{equation} 

Similarly, other forms of WLGI involving three pairs of two-time joint probabilities can be derived by using various combinations of the observable joint probabilities, which are as follows
\begin{subequations}
\label{wlgi2}
\begin{equation}  
\label{}
P(Q_2+,Q_3+) - P(Q_1-,Q_2+) - P(Q_1+,Q_3+) \leq 0   
\end{equation}
\begin{equation} 
P(Q_2+,Q_3-) - P(Q_1-,Q_2+) - P(Q_1+,Q_3-) \leq 0  
\end{equation} 
\begin{equation} 
P(Q_2+,Q_3+) - P(Q_1+,Q_2+) - P(Q_1-,Q_3+) \leq 0   
\end{equation} 
\begin{equation}
P(Q_{1}+,Q_{3}-)-P(Q_{1}+,Q_{2}-)- P(Q_{2}+,Q_{3}-) \leq 0
\end{equation}
\begin{equation}
P(Q_1+,Q_3-) - P(Q_1+,Q_2+) - P(Q_2-,Q_3-) \leq 0    
\end{equation} 
\begin{equation}
P(Q_1+,Q_3+) - P(Q_1+,Q_2+) - P(Q_2-,Q_3+) \leq 0    
\end{equation} 
\begin{equation}
P(Q_1+,Q_3+) - P(Q_1+,Q_2-) - P(Q_2+,Q_3+) \leq 0
\end{equation} 
\begin{equation}    
P(Q_1+,Q_2-) - P(Q_1+,Q_3-) - P(Q_2-,Q_3+) \leq 0    
\end{equation} 
\begin{equation}
P(Q_1+,Q_2-) - P(Q_1+,Q_3+) - P(Q_2-,Q_3-) \leq 0
\end{equation} 
\begin{equation}
P(Q_1+,Q_2+) - P(Q_1+,Q_3+) - P(Q_2+,Q_3-) \leq 0
\end{equation} 
\begin{equation}
P(Q_1+,Q_2+) - P(Q_1+,Q_3-) - P(Q_2+,Q_3+) \leq 0   
\end{equation}

\end{subequations}  

Altering the signs in each of the above Eqs. (\ref{wlgia}) - (\ref{wlgi2}), another set of 12 such 3-term WLGIs can be obtained. 

Now, let the observable $Q$ be measured in $n$ different pairs of instants $t_i$ ($i=1,2,...,n$). From the notion of macro-realism, one can again assume the existence of the overall joint probability distributions $\rho(Q_1,Q_2,...,Q_n)$. Considering the pair-wise observable joint probabilities of the following forms as the marginals of the overall joint probability distributions, we then get the following relation 
\begin{equation}
\label{wig6}
\begin{split}
&P(Q_{1}+,Q_{2}-)+P(Q_{2}+,Q_{3}-)+...+ P(Q_{n-1}+,Q_{n}-) \\& = P(Q_{1}+,Q_{n}-) + (n-2)2^{n-2} \mbox{ non-negative terms}
\end{split}
\end{equation}
From the above expression, the form of WLGI in terms of $n$ pairs of such two-time joint probabilities can be obtained as follows
\begin{equation}
\label{wlgi}
\begin{split}
&P(Q_{1}+,Q_{n}-)- \sum^{n-1}_{i=1} P(Q_{i}+,Q_{i+1}-) \leq 0
\end{split}
\end{equation}
Other various forms of the $n$-term WLGI can be similarly obtained by using different combinations of the joint probabilities for the outcomes $(\pm 1)$ corresponding to $Q_i$'s. However, for illustrating the basic relevant features concerning the efficacy of WLGI, it suffices for our subsequent treatment to confine our attention to essentially 3-term WLGIs involving three pairs of two-time joint probabilities.   \\

Now, considering a typical two-state oscillation, let us focus on a system oscillating between the two states $|A \rangle$ and $|B\rangle$ which are degenerate orthogonal eigenstates of the Hamiltonian $H_0$ corresponding to energy $E_0$, with a perturbing Hamiltonian $H^\prime$ inducing oscillatory transition between these two states, with $\langle A| H^\prime |B\rangle = \langle B| H^\prime |A\rangle = \Delta E$, and $\langle A| H^\prime |A\rangle = \langle B| H^\prime |B\rangle = E^\prime$. Here we take the off-diagonal term of the perturbing Hamiltonian as a real quantity since such a system is typically realized in a double well-potential scenario \cite{ex1,ex8}. The key point here is that at any instant, such a system is found to be either in the state $|A\rangle$ or in the state $|B\rangle$ corresponding to the measurement of the dichotomic observable $Q = |A\rangle \langle A| - |B\rangle \langle B| = P_{+} - P_{-}$ where $P_{+} = |A\rangle \langle A|$, $P_{-} = |B\rangle \langle B|$. Let the  initial state at $t_1$  be of the general form $\rho_0 (t_1) = |\psi_0 \rangle \langle \psi_0|$ where 
\begin{equation}
\label{is}
|\psi_0 \rangle = \cos (\theta) |A \rangle + exp(i\phi) \sin (\theta) |B\rangle
\end{equation} and $\theta \in [0,\pi/2], \phi \in [0,2\pi]$.
 For the above state, the probability of obtaining the measurement outcome, say, +1 at the instant $t_1$ is given by $tr(\rho_0(t_1)P_+)$, and after this measurement, the premeasurement state $\rho_{0}(t_{1})$ changes to the state given by $\rho_{+}(t_1) = P_{+}\rho_0(t_1)P_{+}^{\dagger}/tr(\rho_{0}(t_1)P_+)$ where $P_+ = |A\rangle \langle A| = P_{+}^{\dagger}$. Subsequently, the post-measurement state evolves under the Hamiltonian $H = H_0 + H^\prime$ to the state $\rho_{+}^{\prime} (t_2) = U_{\Delta t}\rho_{+}(t_1)U_{\Delta t}^{\dagger}$ at a later instant $t_2$ where $U_{\Delta t} = exp(-iH \Delta t)$ taking $\hbar = 1$ and $\Delta t = t_2 - t_1$. Then, considering the subsequent measurement of $Q$ at the instant $t_2$, the QM value of, say, the joint probability of obtaining both the outcomes +1 at the instants $t_1$ and $t_2$ is given by
\begin{equation}
\label{eq6}
\begin{split}
& P(Q_1 +, Q_2 +) = tr (\rho_0 (t_1) P_+) tr(\rho_{+}^{\prime} (t_2)P_+) \\& = tr(U_{\Delta t} (P_+ \rho_0 (t_1) P_+) U_{\Delta t}^{\dagger}P_+)
= \cos^2(\theta)\cos^2(\tau)
\end{split}
\end{equation}
where $\tau = \Delta E \Delta t$ (in the units of $\hslash = 1$), and the expression for the unitary matrix $U_{\Delta t}=exp(-iH \Delta t)$ is as follows
\begin{equation}
\label{u}
U_{\Delta t} = e^{-i (E_0+E^\prime) \Delta t} [\cos(\tau) \mathbb I - i \sin(\tau) (|A\rangle \langle B|+|B\rangle \langle A|)]
\end{equation}
Similarly, one can obtain the QM values of the other relevant joint probabilities occurring on the LHS of the 3-term WLGIs given by Eqs. (\ref{wlgia})-(\ref{wlgi2}) (taking $t_2 - t_1 = t_3 - t_2 = \Delta t$).
Consider, for example, the QM expression of the LHS of the inequality (\ref{wlgia}) given by 
\begin{equation}
\label{qm4}
\begin{split}
& P(Q_2+,Q_3-) - P(Q_1+,Q_2+) - P(Q_1-,Q_3-) \\
& = \frac{1}{2}\sin(2\theta) \sin^2(\tau) \sin(2\tau) \sin(\phi)  + \cos^2(\theta)\cos(2\tau)\sin^2(\tau) \\& + \sin^4(\tau) - \cos^2(\theta)\cos^2(\tau) -\sin^2(\theta)\cos^2(2\tau)
\end{split}
\end{equation} which is a function of $\theta,\phi$,and $\tau$, say $f(\theta, \phi,\tau)$. For any given values of $\theta$ and $\tau$, if $\phi$ is varied, it can easily be seen that for $\phi = \pi/2$ (when $\sin(2\tau)$ is positive), or for $\phi = 3\pi/2$ (when $\sin(2\tau)$ is negative), the maximum value of $f(\theta, \phi,\tau)$ is attained. One can thus numerically obtain the simultaneous solution of the following two equations
\begin{widetext}
\begin{equation}
\begin{split}\label{pd}
& \frac{\partial }{\partial \tau} f(\theta, \phi,\tau)_{\phi=\frac{\pi}{2},\frac{3\pi}{2}}= \sin(2\theta)\sin(\phi)_{\phi=\frac{\pi}{2},\frac{3\pi}{2}} (\sin^2(\tau)\cos(2\tau) + \frac{1}{2} \sin^2(2\tau)) + 2\sin(2\tau)\cos^2(\tau) + \sin^2(\theta) \sin(4\tau)=0 \\&
\frac{\partial }{\partial \theta} f(\theta, \phi,\tau)_{\phi=\frac{\pi}{2},\frac{3\pi}{2}}= \cos(2\theta) \sin^2(\tau) \sin(2\tau) \sin(\phi)_{\phi=\frac{\pi}{2},\frac{3\pi}{2}} + \frac{1}{2} \sin(2\theta) \sin^2(2\tau)=0 
\end{split}
\end{equation}
\end{widetext}
to find the solutions for $\theta$ and $\tau$, whence the global maximum of the function $f(\theta, \phi, \tau)$ can be numerically checked. It is then found that for the initial state given by Eq.(\ref{is}) when $\theta = 1.067$ rad, $\phi = \pi/2$ or $3\pi/2$, and $\tau = 1.008$ or 2.133 (in the units of $\hslash = 1$), the maximum QM violation of WLGI given by the inequality (4) occurs with the LHS of Eq. (4) or Eq. (11) given by $f(\theta, \phi,\tau) \approx 0.5043$. It has also been verified that the QM violation of all the 3-term WLGI inequalities (\ref{wlgia})-(\ref{wlgi2}) depend on the the initial state and, among all these inequalities (\ref{wlgia})-(\ref{wlgi2}) and the set of other 3-term inequalities obtained from them, the maximum QM violation is obtained of the inequalities (\ref{wlgia}) and (5a) when the LHS $\approx 0.5043$. In the discussions of the following two sections, we will be specifically considering the form of the 3-term WLGI given by the inequality (4).\\

\section{Wigner's form of LGI and Unsharp Measurement} In the preceding discussion, we have taken the relevant measurements of the observable $Q$ to be essentially `ideal'. Now, if the `non-idealness' of actual measurements is taken into account, a natural question arises as to what effect this would have on the QM violation of WLGI as compared to that for LGI. In order to address this question, we take recourse to the formalism of what is known as unsharp measurement \cite{busch,PB, busch0, spek, busch1} which can be regarded as a particular case of commutating POVM. Note that for an ideal measurement of the dichotomic observable under consideration given by $Q = |A\rangle \langle A| - |B\rangle \langle B|= P_{+} - P_{-}$, the respective probabilities of the outcomes $\pm 1$ and the way a measurement affects the observed state are determined by the projection operators that can be written as $P_{\pm} = (1/2)(\mathbb I\pm Q)$ where $\mathbb I = |A \rangle \langle A| + |B \rangle \langle B|$. Now, in order to  capture the effect of imprecision involved in a non-ideal measurement, using the formalism of unsharp measurement \cite{busch,PB, busch0, spek, busch1}, a parameter ($\lambda$) known as the sharpness parameter is introduced  to characterize the sharpness of a measurement by defining what are referred to as the effect operators given by 
\begin{equation}
\label{eq9}
F_\pm = (1/2)(\mathbb I \pm \lambda Q) = \lambda P_\pm + (1- \lambda) \mathbb I /2 
\end{equation}
\noindent where $(1-\lambda)$ denotes the amount of white noise present in any unsharp measurment $(0<\lambda \leq 1)$, and $F_\pm$ are mutually commuting operators with non-negative eigenvalues; $F_{+}+F_{-} = \mathbb I$, while for $\lambda = 1$ corresponding to sharp measurements, $F_\pm$ reduce to projection operators $P_\pm$. Note that Eq. (\ref{eq9}) can be rewritten as a linear combination of projection operators $P_\pm$ in the following way
\begin{equation}
\label{eq10}
F_\pm = (\frac{1 \pm \lambda}{2}) P_{+}+(\frac{1 \mp \lambda}{2}) P_-
\end{equation}

\noindent Here an important point is that, instead of the projection operators used in the case of an ideal measurement, in an unsharp measurement, the operators $F_\pm$ determine the respective probabilities of the outcomes and the way a premeasurement state changes due to measurement. Considering the generalized $L\ddot{u}ders$ operations, for a specific type of unsharp measurement pertaining to a given state $\rho$, the probability of an outcome, say, +1 is given by $tr(\rho F_{+})$ for which the post-measurement state is given by $(\sqrt{F_{+}}\rho\sqrt{F_{+}}^\dagger)/tr(\rho F_{+})$. Thus, in a given experiment, by estimating the difference between the actually observed probability of an outcome and the corresponding predicted value for an ideal experiment, the sharpness parameter $\lambda$ pertaining to the experiment in question can be determined. This gives an operational significance to the parameter $\lambda$.\\

Using Eq. (\ref{eq9}) or (\ref{eq10}) and by following the prescription outlined above, the QM value of the LHS of WLGI given by (\ref{wlgia}) is now calculated as follows for unsharp measurements pertaining to the two-state oscillation by taking $\rho_0 (t_1) = |\psi_{0}\rangle \langle \psi_{0}|$ where $|\psi_{0}\rangle$ is given by Eq.(\ref{is}), whence one obtains
\begin{widetext}
\begin{equation}
\label{eq11}
\begin{split}
&P(Q_{2}+,Q_{3}-) - P(Q_{1}+,Q_{2}+) - P(Q_{1}-,Q_{3}-) \\& = tr(U_{\Delta t} (\sqrt{F_{+}} (U_{\Delta t} \rho_{0}(t_1)  U^{\dagger}_{\Delta t}) \sqrt{F_{+}}^{\dagger}) U^{\dagger}_{\Delta t} F_{-}) - tr(U_{\Delta t} (\sqrt{F_{+}} \rho_{0}(t_1) \sqrt{F_{+}}^{\dagger}) U^{\dagger}_{\Delta t} F_{+}) - tr(U_{2\Delta t} (\sqrt{F_{-}}  \rho_{0}(t_1) \sqrt{F_{-}}^{\dagger}) U^{\dagger}_{2\Delta t} F_{-}) \\
&= \frac{1}{4} [ 2 \lambda \sin^2 (\tau)(\cos(2\theta) \cos(2\tau) + \sin(2\theta) \sin(2\tau)\sin(\phi) + \cos(2\theta) )  - 2\lambda \sin^2(2\tau) \cos(2\theta)  \\&
 + \lambda \sqrt{1-\lambda^2} \sin(2\tau) (\sin(2\tau) \cos(2\theta) + \cos(2\tau) \sin(2\theta) \sin(\phi)  - \sin(2\theta)\sin(\phi) ) - 2\lambda^2 \cos(2\tau) - \lambda^2 \cos(4 \tau) -1 ]
\end{split}
\end{equation}
\end{widetext}
Considering the situation where the sharpness parameter is unknown, in the above expression, we take certain fixed values of $\theta = 1.0666$ rad, $\phi = \pi/2$ and $\tau = 1.0083$ for the parameters characterizing the initial state and the time evolution. Recall that for these specific choices, as mentioned after Eq. (12) in the preceding section, the QM violation of the 3-term WLGI given by the inequality (\ref{wlgia}) is maximum using the joint probabilities calculated for ideal measurements. Then for these choices of the relevant parameters,  the LHS of WLGI (\ref{wlgia}) or Eq. (15), which is a function of $\lambda$, reduces to the following form
\begin{equation}
\label{un}
\begin{split}
& P(Q_2+,Q_3-) - P(Q_1+,Q_2+) - P(Q_1-,Q_3-) \\& 
= 0.3816\lambda(1-\sqrt{1-\lambda^2})+0.3726\lambda^2 - 0.25
\end{split}
\end{equation}
Note that the above expression (\ref{un}) is a monotonically increasing function of $\lambda \in (0,1]$. This ensures that the solution of
\begin{equation}
0.3816\lambda(1-\sqrt{1-\lambda^2})+0.3726\lambda^2 - 0.25 = 0
\end{equation} provides the critical value of the sharpness parameter $\lambda$ above which, as measurements become more precise, WLGI (\ref{wlgia}) can be violated by the QM predictions. It is then checked that the only solution of Eq. (17) within the allowed range of values of $\lambda$ $(0,1]$ is approximately $ 0.69$.
Thus, within the range $\lambda \leq 0.69$, the maximum QM value of the LHS of (\ref{wlgia})  remains non-positive, implying that within this bound of $\lambda$ for unsharp measurements, the QM predictions always satisfy the 3-term WLGI (\ref{wlgia}).  We shall now compare this critical value of $\lambda$ with that for LGI.

\section{Comparison between WLGI and LGI with respect to Unsharp Measurement}
General form of the LGI involving \textit{n} pairs of two-time correlation functions can be expressed in the following way \cite{review} 
\begin{equation}
\label{eq13}
\begin{split}
& - n\leq K_n \leq n-2 ~~ \text{for odd $n\geq 3$}\\
& -(n-2) \leq K_n\leq n-2 ~~ \text{for even $n\geq 4$ }
\end{split}
\end{equation}
\noindent where $K_n = C_{21} + C_{32} + C_{43} + ..... + C_{n(n-1)} - C_{n1}$, and the correlation function $C_{ij} = \langle Q_{i} Q_{j}\rangle$.
Here we note that for the case of two-state oscillation under consideration, the QM violation of LGI is independent of the initial state \cite{review}, in contrast to the QM violation of WLGI crucially depending on the initial state. For example, for the initial state given by Eq. (8) for which $\theta = 0$; i.e., when the initial state is $|A\rangle$, it can easily be seen that the expression on the RHS of Eq. (11) is essentially a non-positive quantity for any value of $\tau$,  implying that for this initial state, the QM predictions do not contradict the form of WLGI given by the inequality (4). On the other hand, the QM predictions violate LGI in this context, for instance for the initial state $|A\rangle$ that has been explicitly shown \cite{hardy}, thereby indicating the nonequivalence between LGI and WLGI. 

Here it may also be worth mentioning that for an experimental test of the QM violation of the $n$-term WLGI involving the joint probabilities, one has to run the experiment over $n$ sub-ensembles, while for testing the  $n$-term LGI (\ref{eq13}) involving $n$ correlation functions, experiments over $2n$ sub-ensembles are required. 

Next, considering unsharp measurements, the correlation function for any initial state is obtained in the following form 
\begin{equation}
\label{eq14}
\begin{split}
&\langle Q_{i} Q_{j}\rangle_{unsharp} = P(Q_{i}+, Q_{j}+) + P(Q_{i}-, Q_{j}-)\\
& \hskip 2.0cm - P(Q_{i}-, Q_{j}+)- P(Q_{i}+, Q_{j}-)\\
& = tr(U_{\Delta t}(\sqrt{F_{+}}\rho(t_{i})\sqrt{F_{+}}^\dagger)U_{\Delta t}^{\dagger}F_{+})\\
& + tr(U_{\Delta t}(\sqrt{F_{-}}\rho(t_{i})\sqrt{F_{-}}^\dagger)U_{\Delta t}^{\dagger} F_{-})\\
& -tr(U_{\Delta t}(\sqrt{F_{-}} \rho(t_{i})\sqrt{F_{-}}^\dagger)U_{\Delta t}^{\dagger}F_{+}) \\
& - tr(U_{\Delta t} (\sqrt{F_{+}}\rho(t_{i}) \sqrt{F_{+}}^\dagger)U_{\Delta t}^{\dagger} F_{-})\\
& = \lambda^{2} \cos(2\tau) = \lambda^{2} \langle Q_{i}Q_{j}\rangle_{sharp}
\end{split}
\end{equation}

\noindent where $\langle Q_{i}Q_{j}\rangle_{sharp}$ is the correlation function for sharp measurements corresponding to $\lambda = 1$. Using Eq. (\ref{eq14}) and the result that for any given \textit{n}, the maximum QM value of $K_n$ for sharp measurements has been found to be $n \hskip 0.1cm cos(\pi/n)$ \cite{review}, it follows that for unsharp measurements, if the QM predictions are to satisfy the general form of LGI given by (\ref{eq13}), the following inequality needs to hold good 

\begin{subequations}
\begin{equation}
\label{eq15a}
\lambda^{2} n \hskip 0.1cm cos(\pi/n) \leq n-2
\end{equation}
\noindent for any \textit{n}, which implies

\begin{equation}
\label{eq15b}
\lambda \leq \sqrt{\frac{n-2}{n \hskip 0.1cm cos(\pi/n)}}
\end{equation}
\end{subequations}

By evaluating the derivative of the RHS of (20b) with respect to $n$, it is found that as \textit{n} increases ($n\geq 3$), the RHS of (\ref{eq15b}) also increases, thereby implying an increase of the critical value of $\lambda$ (denoted by, say, $\lambda_c$) above which, as measurements become more precise, the QM results can violate the general form of LGI given by (\ref{eq13}). The minimum value of $\lambda_c(=\sqrt{2/3} \approx 0.816)$ occurs for $n=3$. For $n=4$, $\lambda_c$ is given by $(1/2)^{1/4} \approx 0.84$ which is the same as the corresponding $\lambda_c$ \cite{GK} obtained for the Bell-CHSH inequality. Now, comparing the above mentioned minimum value of $\lambda_c (\approx 0.816)$ for the 3-term LGI with the corresponding critical  value of $\lambda_c (\approx 0.69)$ for the 3-term WLGI given by (\ref{wlgia}), it is seen that for the range of values of $\lambda \in (0.69,0.816]$ corresponding to unsharp or imprecise measurements, the QM violation of macrorealism can be tested using the 3-term WLGI, but not in terms of LGI. This, therefore, shows the nonequivalence between WLGI and LGI. 

Here it needs to be stressed that our study is restricted for the choice of a two-level system subjected to a suitable interaction causing oscillation between the two states. Also, to be noted that our study pertains to a restricted class of generalized measurements characterised by the model of unsharp measurements used here that involves commuting POVMs. Of course, the comparative study between WLGI and LGI requires to be extended by considering different types of generalized measurements. Importantly, in view of the recent demonstration of the QM violation of LGI for multilevel systems \cite{BM}, a similar study is required using WLGI in order to establish its efficacy. It should also be worth trying to formulate a comprehensive procedure for constructing and studying the $n$-term WLGIs, analogous to the way higher order LGIs can be constructed from the 3-term case by analyzing the relationship between LGI and the geometry of the `cut polytope' \cite{avis}.

\section{No-Signalling in Time and Unsharp Measurement} As already mentioned in the introductory section, an alternative necessary condition for the validity of macrorealism has recently been proposed \cite{nsit} by assuming that the outcome statistics of a measurement would remain unaffected by any prior measurement. This condition, referred to as `No-Signalling in Time' (NSIT), is the statistical version of NIM used in deriving LGI and can be viewed as the temporal analogue of the no-signalling condition for the spacelike separated measurements used in the EPR-Bohm scenario, with the difference that while any violation of the latter would violate special relativity, the violation of NSIT does not imply any such inconsistency. Now, in order to express NSIT in a mathematical form, let us again consider a system oscillating in time between two possible states. The probability of obtaining a particular outcome, say, +1 for the measurement of a dichotomic observable \textit{Q} at an instant, say, $t_2$, without any earlier measurement being performed, be denoted by $P(Q_{2} = +1)$. NSIT requires that $P(Q_2 = +1)$ should be the \textit{same} even when an earlier measurement of, say, $Q$ is made at an instant, say, $t_1$. In other words, if we denote by $P(Q_2 = +1|Q_1 = \pm 1)$ the probability of obtaining an outcome +1 for the measurement of \textit{Q} at the instant $t_2$ when an earlier measurement of \textit{Q} has been performed at $t_1$ having an outcome $\pm 1$, NSIT can be expressed as the equality condition given by $P(Q_2 = +1) = P(Q_2 = +1|Q_{1}=\pm 1)$ which implies that 
\begin{equation}
\label{eq16}
\begin{split}
P(Q_2 = +1) = P(Q_1 +, Q_2 +) + P(Q_1 -, Q_2 +)
\end{split}
\end{equation}
\noindent where the terms on the RHS of Eq. (\ref{eq16}) are the relevant joint probabilities.\\

Next, pertaining to the two-state oscillation between the states $|A\rangle$ and $|B\rangle$ with the  state $\rho_0(t_1) = |\psi_0\rangle \langle \psi_0|$ at the instant $t_1$ where $|\psi_0 \rangle = \cos \theta |A\rangle + exp(i\phi) \sin\theta |B\rangle$, the QM violation of the condition given by Eq. (\ref{eq16}) for ideal measurements can be obtained as follows, based on calculations similar to that involving Eq. (\ref{eq6}) as discussed earlier
\begin{equation}
\label{eq17}
\begin{split}
&P(Q_2=+1)-[P(Q_{1}+,Q_{2}+)+P(Q_{1}-,Q_{2}+)] \\
& = tr(U_{\Delta t} \rho_0(t_1) U^{\dagger}_{\Delta t} P_{+}) -  tr(U_{\Delta t} (P_{+} \rho_0(t_1) P_{+}) U^{\dagger}_{\Delta t} P_{+}) \\
& - tr(U_{\Delta t} (P_{-} \rho_0(t_1) P_{-}) U^{\dagger}_{\Delta t} P_{+}) \\
& = \frac{1}{2} \sin(2\tau)\sin(2\theta)\sin(\phi)
\end{split}
\end{equation}
It can be seen from Eq. (\ref{eq17}) that, for sharp or ideal measurements, the maximum QM violation of the NSIT condition as given by Eq. (\ref{eq16}) is 1/2 corresponding to the choices $\theta = \pi/4, \phi = \pi/2$ and $\tau = \Delta E \Delta t = \pi/4$ (in the units of $\hslash = 1$). Next, taking into account the unsharpness of measurements involved, the QM violation of the NSIT condition of the form (\ref{eq16}) is obtained as follows on the basis of calculations similar to that leading to Eq. (\ref{eq11})
\begin{equation}
\begin{split}
\label{eq18}
&P(Q_2=+1) - [P(Q_{1}+,Q_{2}+)+P(Q_{1}-,Q_{2}+)] \\&
= tr(U_{\Delta t} \rho_0(t_1) U^{\dagger}_{\Delta t} F_{+}) - tr(U_{\Delta t} (\sqrt{F_{+}} \rho_0(t_1) \sqrt{F_{+}}^\dagger) U^{\dagger}_{\Delta t} F_{+})\\
& - tr(U_{\Delta t} (\sqrt{F_{-}} \rho_0(t_1) \sqrt{F_{-}}^\dagger) U^{\dagger}_{\Delta t} F_{+})\\& = \frac{1}{2} \lambda \sin(2\tau) \sin(2\theta)  \sin(\phi) (1 - \sqrt{1-\lambda^2}) 
\end{split}
\end{equation}
\noindent It is then seen from Eq. (\ref{eq18}) that, while the magnitude of the QM violation of NSIT depends on the value of the sharpness parameter $\lambda$ (this violation is maximum for $\lambda = 1$ corresponding to sharp measurement), a particularly noteworthy feature is that unless the state at the instant $t_1$ is such that either $\sin (2\theta)$ or $\sin (\phi)$ vanishes, the QM violation of NSIT persists for any non-zero value of $\lambda$, \textit{i.e.}, for any arbitrarily unsharp measurement. This shows remarkable robustness of the QM violation of NSIT with respect to unsharp or non-ideal measurements.

\section{Concluding Discussion} Three different necessary conditions for the validity of macrorealism are considered, including the two earlier proposed conditions namely LGI, NSIT, and the alternative condition WLGI proposed in this paper. Comparison between these three conditions in terms of the robustness of their respective QM violations against unsharpness of the relevant measurements is the central theme of the present paper. Our investigation reveals that the QM violation of macrorealism in terms of the macrorealist inequality WLGI can occur for greater imprecision or unsharpness of measurements than that for LGI. Next, coming to NSIT, interestingly, we find that its QM violation occurs whatever be the unsharpness of the relevant measurements. Thus, corresponding to this quantum feature, classicality does not emerge, irrespective of how unsharp the relevant measurements are. Implication of this curious finding calls for further reflection. Here it is relevant to mention that, very recently, Clemente and Kofler \cite{clemente} have proposed a combination of NSIT conditions that are claimed to serve as both necessary and sufficient conditions for macrorealism. It would therefore be interesting to try to extend the investigation of this paper for such a combination of NSIT conditions. We also note the study by Kofler and Brukner \cite{KB} providing an incisive analysis of how classicality emerges within quantum theory for any spin system under the coarse-graining of measurements. Taking into account such works it should, therefore, be instructive to make a comprehensive comparison of the results of our present paper with that using different characterizations of coarse-grained measurements that are invoked while probing the emergence of classicality within quantum theory under imprecise measurements.\\

\section{Acknowledgements.} DH is grateful to Paul Busch for his helpful comments after reading the earlier version of this paper. We thank Swapan Das for necessary help. DH acknowledges support from the Dept. of Science and Technology, Govt. of India and Centre for Science, Kolkata.

\end{document}